\documentclass[ aps,prl,amsmath,amssymb,reprint]{revtex4-1}
\usepackage{color}
\usepackage{graphicx}
\usepackage{dcolumn}
\usepackage{bm}
\usepackage{epstopdf}

\usepackage[caption=false]{subfig}
\usepackage{textcomp}
\usepackage{hyperref}
\usepackage{natbib}

\begin{document}

\preprint{APS-PRL}

\title{Stirring Unmagnetized Plasma}
\author{C. Collins}
\email{cscollins2@wisc.edu}
\author{N. Katz}
\author{J. Wallace}
\author{J. Jara-Almonte}
\author{I. Reese}
\author{E. Zweibel}
\author{C.B. Forest}
\affiliation{Physics Department, University of Wisconsin-Madison}%

\date{\today}

\begin{abstract}
A new concept for spinning unmagnetized plasma is demonstrated experimentally.  Plasma is confined  by an axisymmetric multi-cusp magnetic field and biased cathodes are used to drive currents and impart a torque in the magnetized edge.  Measurements show that flow viscously couples momentum from the magnetized edge (where the plasma viscosity is small) into the unmagnetized core (where the viscosity is large) and that the core rotates as a solid body.  
To be effective, collisional viscosity must overcome the ion-neutral drag due to charge exchange collisions.  
\end{abstract}

\maketitle

Many astrophysical objects---such as most stars and galaxies---are composed of highly-conducting, turbulent, flowing plasma in which the flow energy is much larger than the magnetic-field energy \cite{zweibel97_nature}.  This flow-dominated regime gives rise to well-known instabilities, such as the magnetic dynamo and the magneto-rotational instability \cite{balbus98_rmp}.  These instabilities may give rise to plasma turbulence in which the magnetic field is subdominant to the kinetic energy in the flow.

Thus far, experiments in the flow-dominated regime have used liquid metal instead of plasma.
This is because confinement of hot (conducting) plasma requires strong magnetic fields, and the energy in these fields dominates the energy in the flow.  Nevertheless, although confinement of liquid-metal is straightforward, the medium suffers from several limitations.  Liquid metals cannot capture plasma effects beyond magnetohydrodynamics (MHD), and their magnetic Reynolds number $R_{\rm m}=\mu_0\sigma V L$, which governs self-excitation by magnetic fields, is limited by achievable flow speed $V$ ($L$ and $\sigma$ are respectively characteristic length and conductivity).  Furthermore, the product of conductivity and viscosity, the magnetic Prandtl number $P_{\rm m}=\mu_0\sigma\nu$, is fixed in liquid metals at around $10^{-5}$. 
In comparison, plasma flow speeds can be fast (km/s), resulting in large $R_{\rm m}$; moreover, plasma viscosity can be varied independently of the conductivity allowing $P_{\rm m}$ to vary from the liquid-metal/stellar interior regime $(P_{\rm m}\ll$ 1) to the regime of interstellar plasmas ($P_{\rm m}\gg$1). Experimental control of $P_{\rm m}$ is important for understanding the nature of plasma turbulence and the dissipation range.

Therefore, we seek to create a large, steady-state, fast flowing, hot plasma which is weakly magnetized; this combination of parameters is common for astrophysical and space plasmas but has never been studied in a laboratory plasma. Recently, a technique has been proposed for stirring an unmagnetized plasma  \cite{spence09_apj}. The concept relies on axisymmetric rings of alternating-polarity permanent magnets to provide confinement on the boundary; the high-order multipolar magnetic field drops off quickly with distance from the wall so that the core is unmagnetized. The edge of the plasma is then stirred by applying electric field across the magnetic field, and it is expected that viscosity will couple the flow inward to the unmagnetized region. In part, the proposed scheme was motivated by electrode biasing experiments that have been used to control rotation in many magnetized plasma configurations, including mirrors~\cite{Ghosh2006_pop}, tokamaks~\cite{taylor89_prl}, reversed field pinches~\cite{almagri98_pop, craig97_prl}, plasma centrifuges\cite{krishnan81_prl} and other devices \cite{lehnert71_nf}.    

In this Letter, we describe the experimental realization of a rotating, unmagnetized plasma in a novel Couette geometry.  Our measurements of azimuthal velocity show that flow induced in the magnetized region viscously couples to the unmagnetized bulk plasma. The measured velocity profiles also provide for the first time a direct measurement of the unmagnetized plasma viscosity.  The experiment may therefore be used as a viscometer to measure the anisotropic plasma viscosity and compare it to theory.  This modest experiment has already achieved $R_{\rm m}\sim25$ and $P_{\rm m}\sim 0.3-6$,  which is approaching regimes shown to excite the magneto-rotational instability in local linear analysis and global Hall-MHD numerical simulations \cite{ebrahimi2011_pop}.

\begin{figure*}[t]
\subfloat[]{\label{fig:chambera}\includegraphics[height=2.75in]{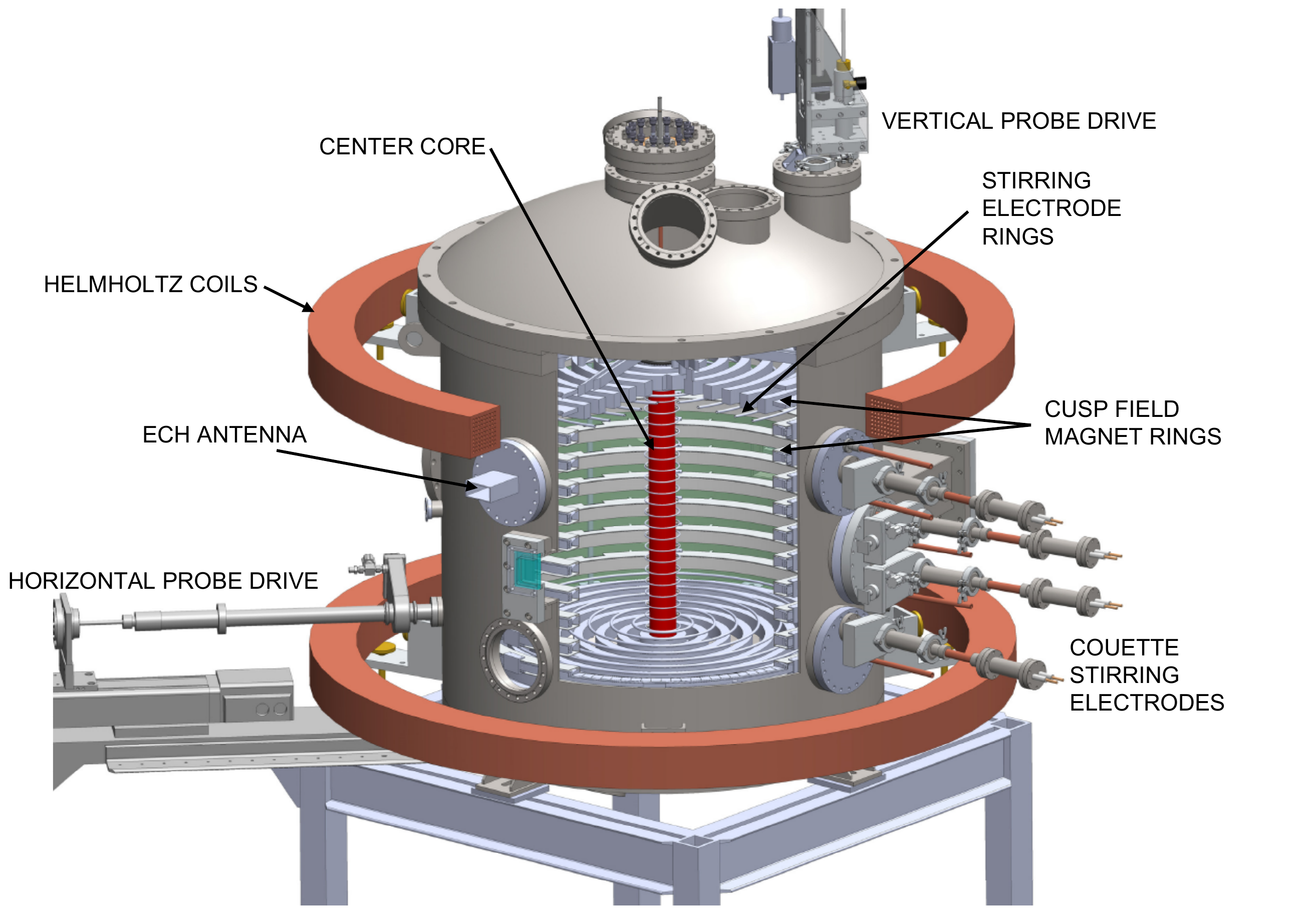}}
\subfloat[]{\label{fig:chamberb}\includegraphics[height=2.75in]{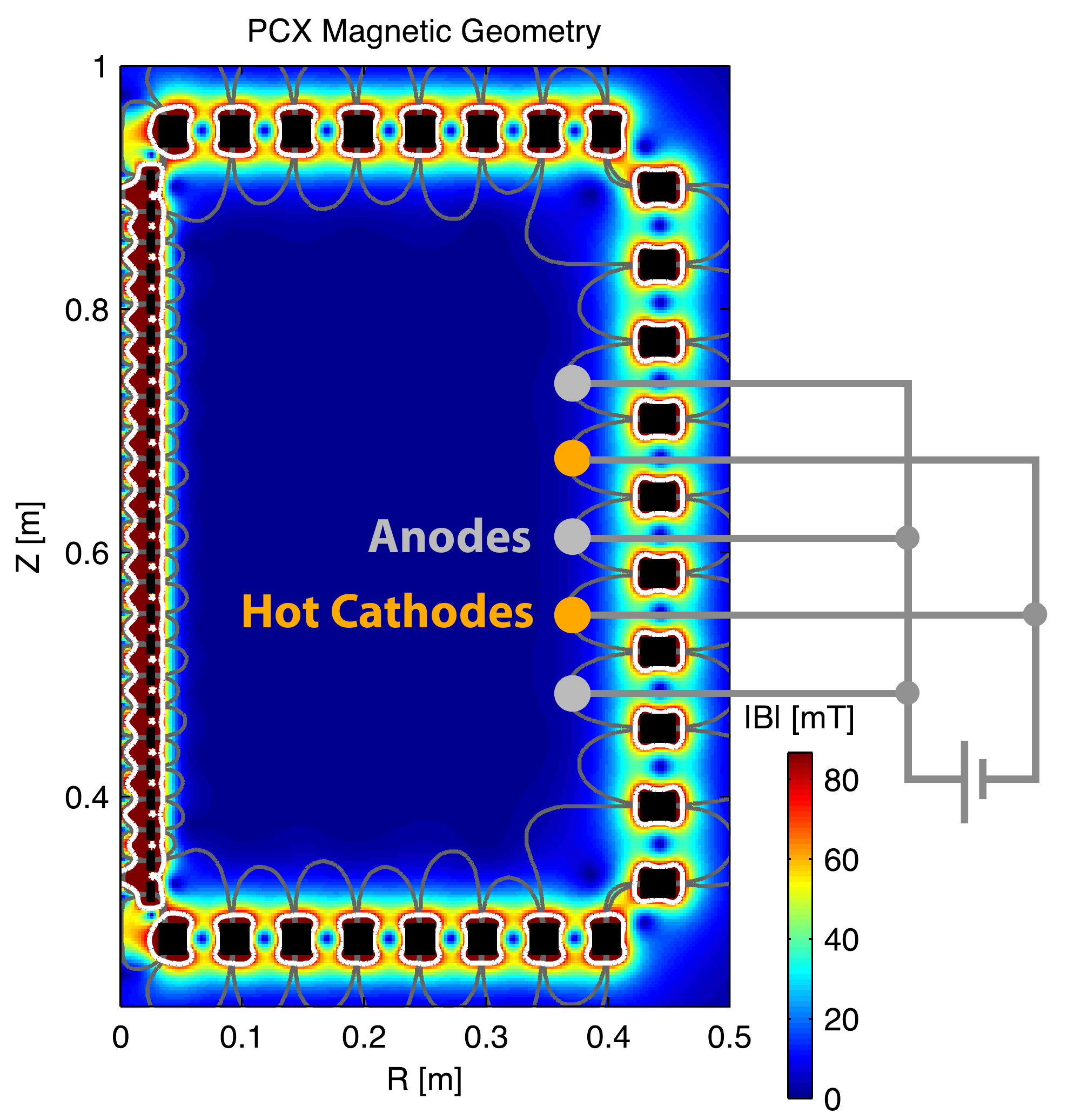}}
\caption{\label{fig:chamber}(color online).The Plasma Couette Experiment (PCX). (a) Experimental setup showing the vacuum vessel, internal structure of the ring cusps and diagnostic layout; (b) Poloidal cross-section showing magnetic geometry, and stirring electrodes; black rectangles are permanent magnets; white lines show resonance surfaces for electron-cyclotron heating ($|B|=87$ mT); gray lines show cusp field lines. } 
\end{figure*}

\begin{figure*}
\subfloat[]{\label{fig:parama}\includegraphics[width=.45\textwidth]{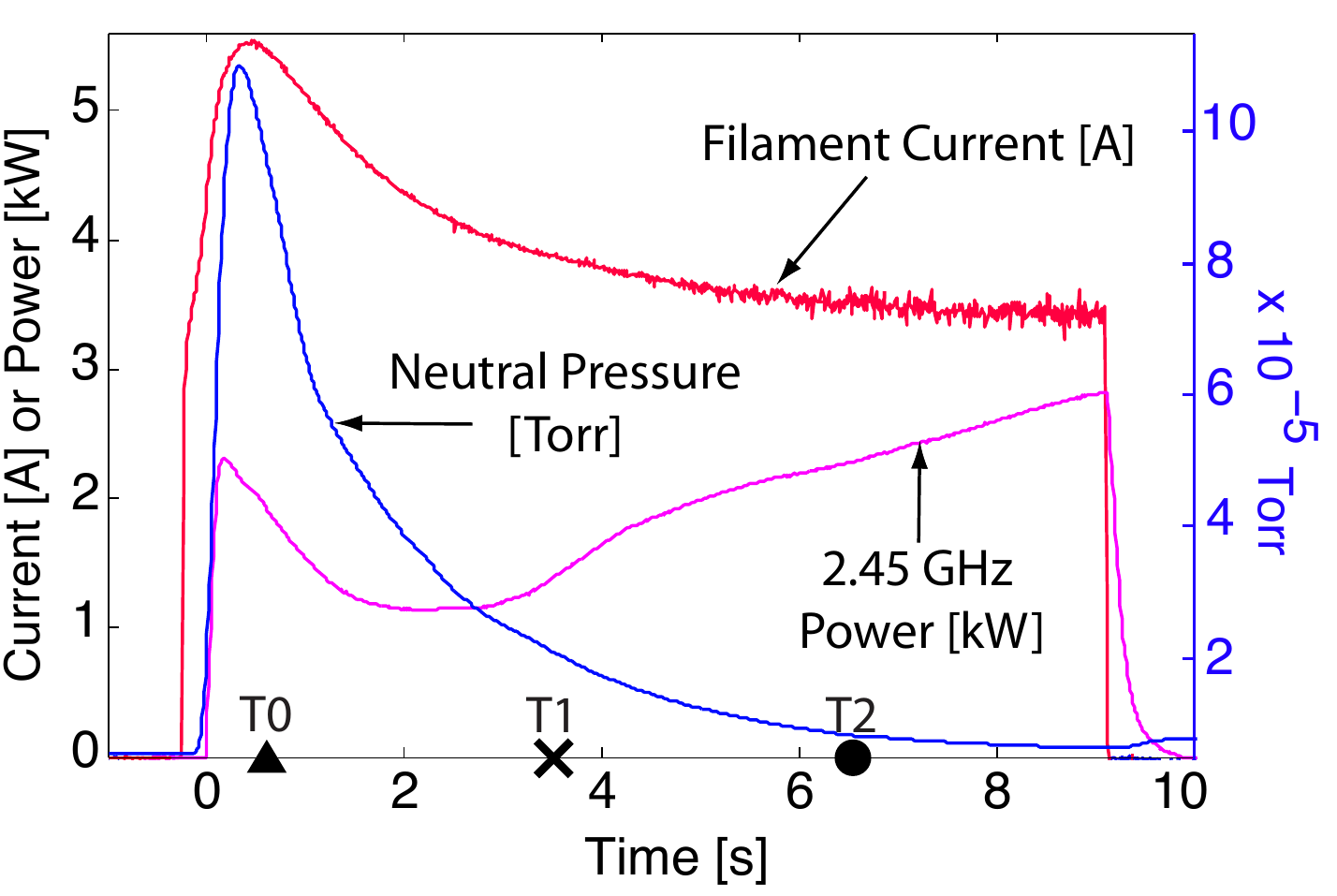}}
\subfloat[]{\label{fig:paramb}\includegraphics[width=.45\textwidth]{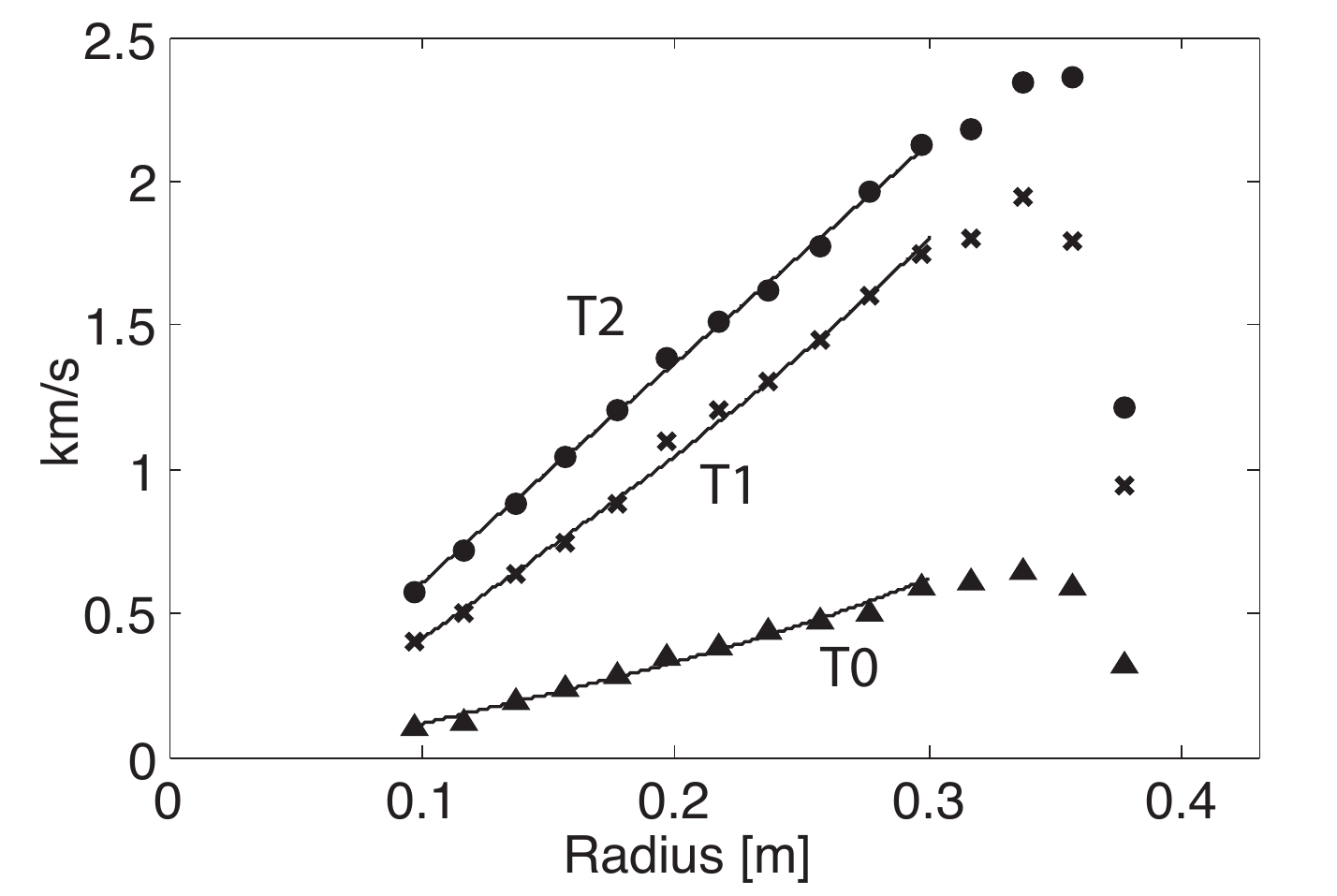}}

\caption{\label{fig:param}  a) Typical plasma discharge: plasma density decreases and electron temperature increases as the puff of neutral gas is pumped out. b) Velocity profiles in argon at times labeled in (a) for two cathodes biased to 400 V. Rotation increases at low density and high ionization fraction. Solid lines represent a model incorporating unmagnetized Braginskii viscosity and neutral drag (due to charge exchange), using parameters from Table \ref{tab:velocitytable}.  }
\end{figure*}

The Plasma Couette Experiment setup is shown in Fig.~\ref{fig:chambera}.  Confinement is provided by a cylindrical assembly of permanent magnets, arranged in rings of alternating polarity, to form an axisymmetric cusp magnetic field. A center core (red) of small ring magnets provides the inner boundary.   As shown in Fig. ~\ref{fig:chamberb}, there is a large region (0.6 m diameter, 0.6 m tall) of unmagnetized plasma between the inner and outer cusp boundaries. The plasma is created by 2.45 GHz electron-cyclotron heating in the magnetized edge. Even though the resonance zone is very close to the face of the magnets, the heating and confinement are sufficient to create plasmas with densities near the cutoff ($7\times10^{10}$ cm$^{-3}$), and electron temperatures as high as 12 eV (measured by a single-tip swept Langmuir probe).  Consistent breakdown requires a strong puff of gas at the beginning of each shot (10$^{-4}$ torr for helium and 10$^{-5}$ torr for argon).  As the gas is pumped out (see Fig.~\ref{fig:parama}), the plasma density decreases and the electron temperature increases. Thus a wide range of plasma parameters can be studied over the course of a single shot. The shot length is limited to about 10 seconds by magnet cooling requirements.  Note that although the plasma flow is changing throughout the shot, we have quasi-steady state because the viscous time $L^2/\nu\sim10^{-4}$ s is short.

Flow is measured using a Mach probe, in which two oppositely-facing molybdenum disks separated by boron nitride insulator are biased to draw ion saturation current. Drift velocity can be deduced from the ratio of upstream to downstream current ($j^+/j^-=\mathrm{exp}(K v_{d}/c_s)$, with K=1.34 for $T_i < 3T_e$ \cite{hutchinson02}). Unmagnetized Mach probes have been shown to be reliable when the Debye length $\lambda_{De}$ is small relative to the probe size $R_p$ \cite{hutchinson03,ko06}, which is the case in this experiment where $\lambda_{De}\le.04 R_p$.  The probe position is changed from shot-to-shot to measure the radial profile.

The plasma is stirred by ${\bf J \times B}$ torques, with current driven by electrostatically-biased electrodes in the magnetized edge region.  Two sets of electrodes are tested.  The first set consists of toroidally continuous rings of molybdenum-coated copper tubes, inserted between the top and bottom magnet rings and biased with alternating polarity (see Stirring Electrode Rings in Fig.~\ref{fig:chambera}).  Note that the  surface of the magnets is electrically insulated with a ceramic; this insulation prevents currents from shorting out the  cross-field electrostatic potential established by  the electrodes.  Rotation velocities up to 3 km/s are achieved in narrow regions at the corners of the experiment; furthermore, rotation can be reversed by changing the polarity of the applied electric field, as expected.  Significant improvement in the rotation drive is achieved with the second set of electrodes, which consists of thermionic cathodes at a fixed toroidal location (see Couette Stirring Electrodes in Fig. ~\ref{fig:chambera}). The negatively-biased, electron-emitting cathodes draw 10 times the current of the first set of electrodes, despite having $10^5$ smaller surface area.  This smaller area also reduces the particle and energy fluxes escaping the plasma. Each cathode consists of one or more strands of 0.015 in. thoriated tungsten heated to temperatures above 1700$^{\circ}$C. The cathodes are biased 400-500 V with respect to cold anodes in adjacent private flux regions, and the resulting cross-field current is likely ion Pedersen current.  The entire circuit floats electrically with respect to the plasma and the vessel wall. At present, we are only using two cathodes and three anodes, as shown in Fig.~\ref{fig:chamberb}, and torque is only applied at the outer boundary.

Velocity profiles for argon are shown in Fig.~\ref{fig:paramb} for three different times, and therefore three different neutral pressures.  As the neutral gas is pumped out, the rotation velocity increases everywhere with time. In the bulk, unmagnetized plasma, the velocity is approximately linear with radius, implying solid-body rotation.

Although the rotation drive is toroidally localized, the Mach probe successfully measures rotation on the opposite toroidal side of the chamber (see Horizontal Probe Drive, Fig.~\ref{fig:chambera}), suggesting that flow is axisymmetric. Isolated measurements with Mach probes at other toroidal locations are consistent with axisymmetric flow to within the accuracy of the measurements. 
The reason for the symmetry involves the $\nabla B$ and curvature drifts, and the ${\bf E \times B}$ flows themselves:  these drifts are all in the toroidal direction and act to symmetrize the plasma potential at the edge.  Thus, even though the current is only injected at one toroidal position, the electric field at the edge becomes axisymmetric.  

For toroidal velocity to couple inward from the edge and stir the unmagnetized core plasma, the viscous momentum diffusion must be greater than the drag due to ion collisions with neutrals.  This requirement can be deduced from the toroidal momentum equation
\begin{equation}
mn\frac{dv_\phi}{dt}= ({\bf J}\times{\bf B})_\phi - \frac{mn v_\phi}{\tau_{i0} }+ mn\nu[\nabla^2\mathbf{v}]_\phi,
\label{momentum}
\end{equation}
where $m$ is the ion mass, $\tau_{i0}  \approx (n_0 \langle\sigma_{cx} v\rangle)^{-1} $ is the ion-neutral collision time, $\sigma_{cx}$ is the charge-exchange cross-section, and $\nu$ is the viscosity due to ion-ion collisions.

The first term on the right-hand side is the electromagnetic force on the plasma from the injected current in the magnetized region. This term vanishes in the core of the plasma where the magnetic field is essentially zero. The second term is the momentum loss due to charge-exchange collisions with neutrals. Since ionization mean-free path for neutrals is longer than experiment size, neutrals are not strongly coupled to ions and we can assume a uniform background of stationary neutrals.  The final term couples momentum inward through viscosity.  

We turn now to the flow in the unmagnetized region. In steady state with ${\bf v}=v_\phi{\bf e}_\phi$, the left-hand side of Eq.\eqref{momentum} is zero. Also, in the unmagnetized region, we have ${\bf J}\times{\bf B}=0$.  If we further assume that the system is uniform in the vertical direction, the velocity in the bulk is determined by 
\begin{equation}
\frac{\partial^2 v_\phi}{\partial r^2} + \frac{1}{r} \frac{\partial v_\phi}{\partial r}  - \left( \frac{1}{L_v^2} + \frac{1}{r^2} \right) v_\phi = 0
\end{equation}
where $L_v^2\equiv \tau_{i0} \nu$. The solution involves modified Bessel functions
\begin{equation}
v_\phi(r) = A {\rm I}_1(r/L_v )+ B  {\rm K}_1(r/L_v ),
\label{vphi}
\end{equation}
where the constants $A$ and $B$ depend on the velocity applied at the outer and inner boundaries of the unmagnetized region. Note that in the absence of ion-neutral collisions, the solution is simply Taylor-Couette flow with $v_\phi\propto 1/r$.

In order for the unmagnetized plasma to spin we require that the characteristic momentum diffusion length scale, $L_v$, be comparable to the experiment size or larger.  When $L_v$ is small, the rotation (Eq.\eqref{vphi}) remains confined near the edge region.  We may write out the dependence of $L_v$ on plasma parameters by using Braginskii's unmagnetized viscosity $\nu=0.96v_{ti}^2\tau_{i}$ \cite{braginskii65_rpp}:
\begin{equation}
L_v  = \sqrt{\tau_{i0}\nu}= \mbox{21  cm} \frac{\sqrt{f_{i,1\%}} T_{i,eV} }{n_{11}},
\end{equation}
where $f_{i,1\%}$ is the ionization fraction in percent ($f_{i}\equiv n/n_0$), $T_{i,eV}$ is the ion temperature in eV and 
$n_{11}$ is the electron density in units of $10^{11}$ cm$^{-3}$. Therefore, velocity couples into the core for low density, high ion temperature and high ionization fraction, or in other words, when the viscous momentum diffusion is large compared to the ion-neutral drag. Note that Braginskii's viscosity only applies when the ion-ion collision time satisfies $\tau_{ii}\ll\tau_{i0}$.

From the measured velocity profiles and Eq.\eqref{vphi} we may find $L_v$ and thereby measure the ion temperature in the unmagnetized region.  For this measurement we use the plasma and neutral densities and the charge-exchange cross-section.   For argon \cite{hegerberg82}, we take $\sigma_{cx}=6\times10^{-19}T_i^{-0.093}$ m$^2$, while for helium \cite{helm77} we have $\sigma_{cx}=3\times10^{-19}T_i^{-0.106}$ m$^2$.  Since $v_{ti}>v_\phi$, we approximate $\langle\sigma_{cx}v\rangle\approx\sigma_{cx}v_{ti}$.  The only fit parameter is the ion temperature.  The fits for argon, shown in Fig.~\ref{fig:paramb}, use parameters from Table \ref{tab:velocitytable} and demonstrate excellent agreement with the velocity data.

\begin{figure}
\subfloat[]{\label{fig:magnetizationa}\includegraphics[width=.45\textwidth]{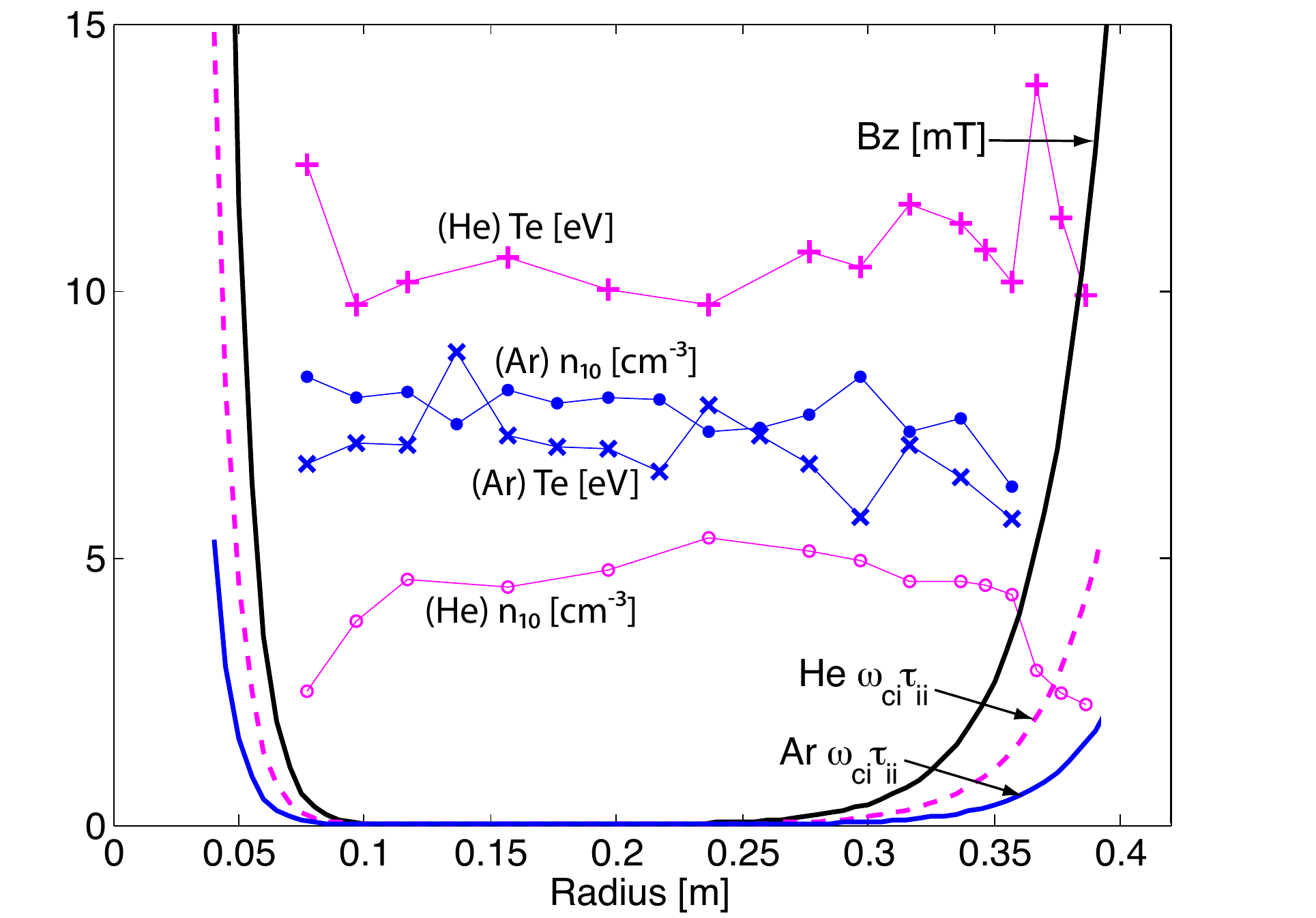}}\\
\subfloat[]{\label{fig:magnetizationb}\includegraphics[width=.45\textwidth]{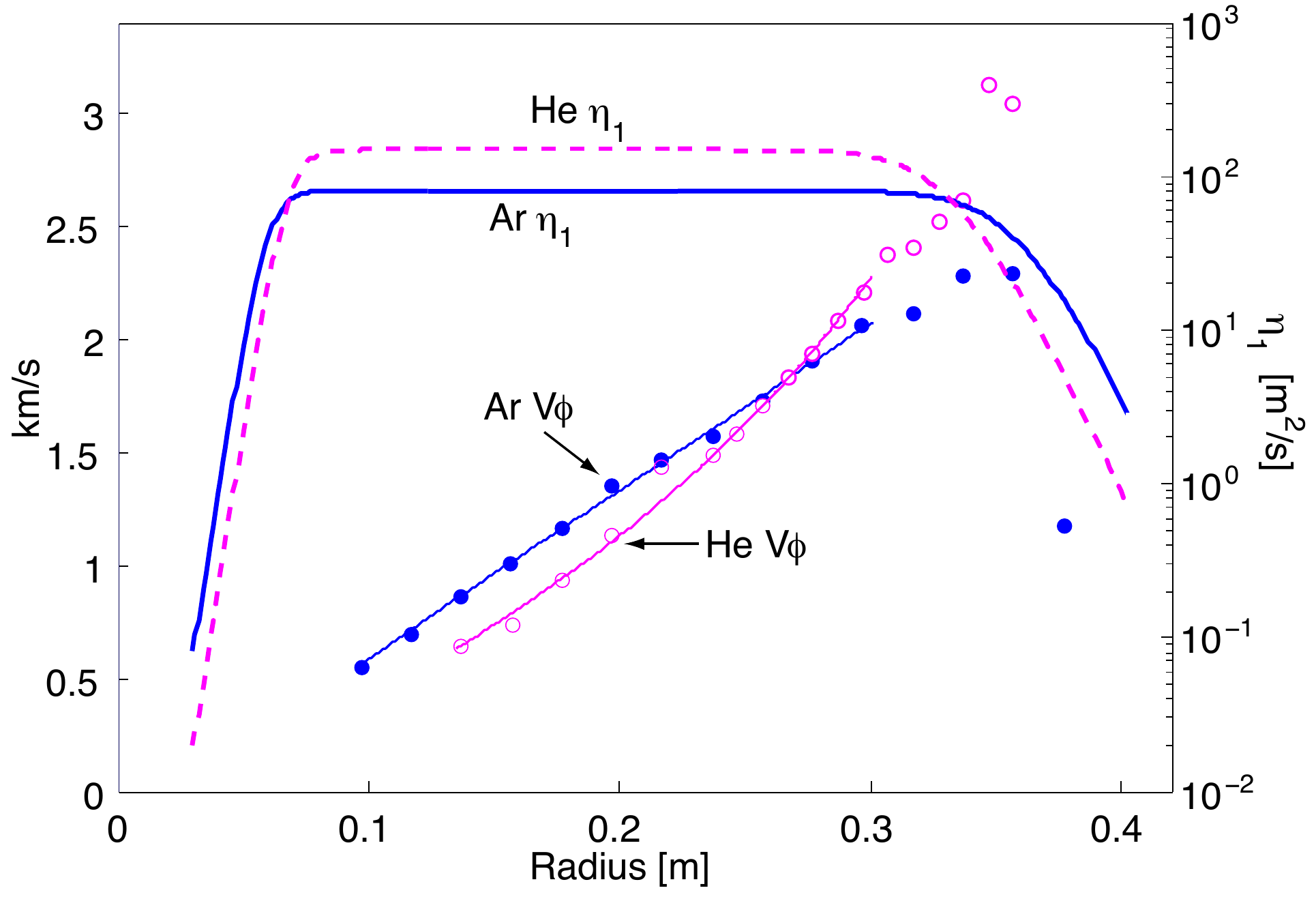}}
\caption{\label{fig:magnetization}  a) Typical density and temperature measurements in argon and helium, along with a model of magnetic field profile and magnetization parameter $\omega_{ci}\tau_{ii}$ . Ions are magnetized when $\omega_{ci}\tau_{ii}>1$. b) Viscosity ($\eta_1$) \cite{braginskii65_rpp} varies continuously from unmagnetized to magnetized regions; torque is applied at $r\approx0.36$ m to induce rotation at the outer boundary.  The calculated viscosity assumes constant density and temperature values as listed under ``He400VT1'' and ``Ar400VT2'' in Table \ref{tab:velocitytable}. }
\end{figure}

Alternatively, we may assume $T_i\sim0.3-0.6$ eV (typical of plasmas confined in multicusps \cite{hershkowitz75_prl}) to compute $\tau_{i0}$ and then we have a measurement of viscosity.  In multicusp plasmas, the viscosity changes dramatically over the plasma profile:  in the edge where the plasma is magnetized, the cross-field kinematic viscosity is $\nu\propto \rho_i^2 / \tau_{ii}$, where $\tau_{ii}$ is the ion-ion collision time and $\rho_i$ the ion gyroradius; in the core, the plasma is unmagnetized and the viscosity increases dramatically to  $\nu \propto v_{ti}^2\tau_{ii}$,
 as shown in Fig. \ref{fig:magnetizationb}. The viscosity can be expressed for arbitrary magnetization according to $\eta_1$ in \citet{braginskii65_rpp}.

The demonstration of plasma flow in the unmagnetized core region of a multicusp device is a major step towards future flow-dominated experiments at parameters relevant to astrophysical plasmas. In the near term we hope to implement the full Couette geometry with differentially rotating plasma by adding cathodes to the inner boundary, and to
attempt von K\'arm\'an flow by rotating the top and bottom in different directions. The concept also lends itself to being a simple plasma centrifuge. 
Two shortcomings of these results are the lack of a reliable ion temperature measurement for detailed comparisons between Braginskii's formulas for viscosity and the measured viscosity, and the reliance upon Mach probes for flow measurements.  These results should be confirmed by using laser-induced fluorescence to measure both the ion flow and the ion temperature.  
Furthermore, if the flow has axial dependence, then there would likely be poloidal flow; this important issue will be addressed in more detail in future publications.  We would also like to better understand the nature of the boundary condition for flow, both in the regions where the plasma is being stirred and in the regions where there are no electrodes.  Our preliminary measurements in the core suggest a free-slip boundary condition at the top and bottom of the cylinder.

\begin{acknowledgments} 

This work was funded in part by NSF and DOE. C.C. acknowledges support by the ORISE Fusion Energy Sciences fellowship, and N.K. acknowledges support from the DOE postdoctoral fellowship.
\end{acknowledgments}

\begin{table}\small 
\begin{tabular}{lcccccccccc}
	\hline
	\\[1pt]
$$\bf{Case}  &  \bf{n} x 10$^{11}$  & $f_{ion}$ &  \bf{Te}  &\bf{Ti}$_{fit}$ &  \bf{V$_{max}$}  & \bf{Rm}  & \bf{Re}&  \bf{Pm} \\
       &(cm$^{-3}$) & $\%$  & (eV)  &(eV) &(km/s) &        &         &       &    \\            	
	\hline
He500VT1   & .47 & 0.7	& 10    & 0.35     &  3  &  23  & 8&2.7 \\
He400VT1   & .45 & 0.6	& 11    & 0.4     & 2.4&  22  & 5  & 4.5\\
He500VT0   & 1.2 & 0.5	& 6     & 0.8       & 1.6 &  6  & 1   & 4.7\\
He400VT0   & 1.2 & 0.4	&  6     & 0.85         & 1.1 &  4 & 0.8   &5.5\\
Ar400VT2   & .77 &  27	& 7     & 0.6          & 2.2 & 11& 8 & 1.3\\
Ar400VT1   & 1.1 &  15	 & 5.9  & 0.42         & 1.8 & 7 & 23 & 0.3\\
Ar400VT0    & 4.2 & 12	  & 2.6  & 1.2       & 0.6 & 0.8       &2 & 0.5 \\
	\hline
\end{tabular}
\caption{Summary of maximum induced velocities for different gas species (helium, argon), cathode voltages (400-500 V), and times  throughout a plasma pulse (T0,T1,T2 as in Fig.~\ref{fig:param}); these times are characterized by varying plasma densities, ionization fractions, and temperatures.  }
\label{tab:velocitytable}
\end{table}

\bibliography{stirringbib}

\end{document}